\documentclass[12pt]{article}
\usepackage{graphicx}
\usepackage{enumerate}
\usepackage{amsfonts}

\newtheorem{remark}{Remark}
\usepackage{subfigure}
\newcommand{\includegraph}[2][]{\ifnum\pdfoutput=0\includegraphics[#1]{#2.eps}\else\includegraphics[#1]{#2.pdf}\fi}
   \usepackage[english]{babel}

\begin{document}

\title{Mixed mode oscillations in the Bonhoeffer-van der Pol  oscillator with weak periodic perturbation}
\author{Ekaterina Kutafina\\
Hasselt University, Agoralaan–-Gebouw D,\\
 3590 Diepenbeek, Belgium\\
 and\\
   AGH University of Science and Technology,\\
 al. Mickiewicza 30 30-059 Krakow, Poland. \\
Email: {\it ekaterina.kutafina@gmail.com}}
 
\maketitle

\begin{abstract} Following the paper of K. Shimizu et al. (2011) we consider the Bonhoeffer–-van der Pol  oscillator with non-autonomous  periodic perturbation. We show that the presence of mixed mode oscillations reported in that paper can be explained using the geometrical theory of singular perturbations. The considered model can be re-written as a 4-dimensional (locally 3-dimensional) autonomous system, which under certain conditions has a folded saddle-node singularity and additionally can be treated as a three time scale one. \end{abstract}
 \section{Introduction}

 There exist a lot of different mechanisms which can create a mixed mode oscillations (MMOs) in multiple time scale system of differential equations. In the paper of Desroches et al. (2012) one may find an information about  MMOs near a singular Hopf bifurcation, near a folded node etc. analyzed  from the perspective of singular perturbation theory and blow up methods (see the book written by F. Dumortier (1993) for theoretical backgrounds). Most of the research in this area is made for autonomous systems, but there exist some results for non-autonomous systems as well (e.g. Szmolyan and Wechselberger 2004).  
 In what follows we focus on the system presented in  the paper of Shimizu et al. (2011)
  \begin{equation}\label{fvdp}
        \left\{
        \begin{array}{ll}
        \epsilon\dot{x}=y+x-x^3\\
        \dot{y}=-x-k_1\,y+B_0+B_1\,\sin\omega\,t.\\
        \end{array}\right.
 \end{equation}
This system could be seen as one of the versions of the forced van der Pol oscillator. The great advantage of this model is that it represents a simple electric circuit, so experimental results are easily available. The small parameter $\epsilon>0$ is responsible for a singular perturbation, while $B_1$ is also assumed to be small and positive and controls a regular non-autonomous periodic perturbation. For $B_1=0$ system (\ref{fvdp}) is a two dimensional autonomous system with classical canard explosion (see Krupa and Szmolyan 2001) controlled by the parameter $B_0$. The second parameter $k_1$ makes enough room for a generalized  Hopf bifurcation  (see the book of Yu. Kuznetsov 2004) in the planar system. This bifurcation (also known as Bautin bifurcation) takes place when classical Hopf bifurcation degenerates due to the vanishing first Lyapunov index.  Using  MATCONT, a continuation package for Matlab (see Dhooge et al. 2003) we may easily compute that Bautin bifurcation happens at $k_1\approx 0.513$ and $B_0\approx 0.365$.   The diagram presented by Shimizu et al. (2011) shows that for $k_1=0.9$ a subcritical Hopf bifurcation is observed for $B_0\approx 0.20543$.  Additional computations made for $k_1=0.2$ detect a supercritical bifurcation at  $B_0\approx 0.4945$. These samples will be  useful for the numerical computations presented in the next sections. 

Systems very similar to (\ref{fvdp})  were widely studied since almost the beginning of the previous century (see van der Pol 1920). In the papers by Guckenheimer et al. (2003) and Bold et al. (2003) the results of  the very extended bifurcation analysis are presented together with the references to many works on the topic.

The full  system (\ref{fvdp}) reveals  a very rich dynamics, for instance in the paper by Shimizu et al. (2011)
 authors present a sequence of stable mixed mode oscillations, which seems to be different from standard Farey sequence reported e.g. by
Petrov et al. (1992). The key  difference is the fact that in  the system (\ref{fvdp}) we can easily find a periodic solution with more than two large oscillations in a row. In fact, it seems to be very difficult to find a MMO with only one large oscillation. Moreover the regions of parameters where the periodic MMOs exist are relatively small. We may also observe chaotic oscillations and bursting phenomena. 

The theoretical background  for  multi-scale non-autonomous systems is not yet very  developed. One of the typical approaches is to consider a time variable as an independent one (see Szmolyan and Wechselberger 2004). Then we can apply the theory valid for autonomous system and get better intuition about the dynamics.
In this paper we also  propose to rewrite the system as an autonomous one, but (it may seem to be artificial on the beginning) we decide to introduce two new variables $z=\sin\omega\,t$ and $p=\cos\omega\,t$. This approach allows us to reveal better analogy to the known tree-dimensional systems with a fold singularity
and explain the presence of MMOs  using the methods known from geometric singular perturbation theory. 

  We  write down a four dimensional system, keeping in mind an additional equation $z^2+p^2=1$, which makes it in fact a locally three dimensional:
 \begin{equation}\label{4D}
        \left\{
        \begin{array}{ll}
        \epsilon\dot{x}=y+x-x^3\\
        \dot{y}=-x-k_1\,y+B_0+B_1\,z\\
        \dot{p}=-\omega\, z\\
        \dot{z}=\omega\, p.
        \end{array}\right.
 \end{equation}
 	\begin{remark} For the further analysis it is convenient to see the system (\ref{4D}) as a three dimensional system in the space $(x,y,z)$, but due to the trigonometrical origin of $z$  there are two different $p$ values in each fixed point, thus two different values of $\dot{z}$ (difference is only in the sign). That means that at each fixed $(x,y,z)$ there intersect two trajectories - one following  increasing values of $z$ and another one following decreasing values. 
On the figures it is sometimes better to show the trajectories in $(x,y,p)$ space, the small oscillations are better visible this way. \end{remark}
 	
For $\epsilon=0$ system (\ref{4D}) has a slow manifold  given by a surface $y=-x+x^3$ which has folds at $x=\pm\frac{\sqrt{3}}{3}$. Fenichel's theory (see Jones et al. 1995 for details) covers the dynamics of the perturbed system near normal hyperbolic points of the slow manifold. In case of the system (\ref{4D}) normal hyperbolicity condition fails near the folds. 
It occurs that the dynamics reported by Shimizu et al. (2011) shows many similarities to the prototype three-scale model with a folded saddle-node singularity of the type II presented by Krupa et al. (2008):
\begin{equation}\label{KP}
        \left\{
        \begin{array}{ll}
        \epsilon\dot{v}=-z+f_2 v^2+f_3 v^3\\
        \dot{z}=v-w\\
        \dot{w}=\epsilon(\mu-g_1 z).
        \end{array}\right.
 \end{equation}
 The generalized Bonhoeffer–-van der Pol system investigated by Sekikawa et al. (2010)
\begin{equation}\label{jap}
        \left\{
                \begin{array}{ll}
        \epsilon\dot{x} = x(1-x^2)+y+z    \\
                    \dot{y} = -x-k_1\,y+B_0   \\
                    \dot{z} = k_3(-x-k_1\,z+B_0)
        \end{array}\right.
 \end{equation}
 is in fact even  closer to (\ref{fvdp}) and it was shown  by De Maesschalck et al. (preprint) that under certain conditions it can be treated as a three time scale system.
In this paper we skip the full formal analysis of the system (\ref{4D}). Instead our intention is to show what parts of the investigation could be done analogously to the  papers of  Krupa et al. 2008  or De Maesschalck et al. and what is different and requires a new approach. We assume that the parameters are not essentially different from those used in the paper of Shimizu et al. (2011), otherwise the behavior of the system could change drastically (see the Discussion).

In the following sections we start with a change of  coordinates which  move the right fold of the slow manifold to the origin and brings the system (\ref{4D}) as close as possible to the form of (\ref{KP}). Then we project the flow on a slow manifold and check what kind of folded singularities are present. We discover that for certain values of bifurcation parameter $\mu$ the system can have a folded saddle-node. Rescaling of the coordinates makes the dynamics near the fold more  clear. We show, that analogously to the results of De Maesschalck et al. (preprint) it is possible to consider the flow near the fold as a perturbation of the certain integrable system. Skipping the other blow-up charts (or exit and entrance areas, according to Krupa et al. (2008)) we make some calculations to approximate a return mechanism.
Our analysis is essentially based on the theory built by Krupa et al. (2008), so in order to keep our paper short we do not repeat the information given there.
  
\section{Transition to a standard form}    
It is convenient to transform the variables of the system (\ref{4D}) in the way that the right fold of the slow manifold is moved to the origin and is placed in a local minimum (see De Maesschalck et al. (preprint)).
 Thus we make a following change of coordinates:
 \begin{equation}x=-X+\frac{\sqrt{3}}{3},\qquad y=Y-\frac{2}{9}\sqrt{3},\qquad p=\frac{P}{B_1},\end{equation}
$$  z=\frac{Z}{B_1}+\frac{3\sqrt{3}-2k_1\sqrt{3}-9B_0}{9B_1}. $$
 If we additionally introduce a new parameter $\mu=-\frac{3\sqrt{3}-2k_1\sqrt{3}-9B_0}{9}$ then (\ref{4D})
 takes form
 \begin{equation}\label{4D1}
        \left\{
        \begin{array}{ll}
        \epsilon\dot{X}=-Y+\sqrt{3}X^2-X^3\\
        \dot{Y}=X-k_1\,Y+Z\\
        \dot{P}=\omega\, (\mu -Z )\\
        \dot{Z}=\omega\, P.
        \end{array}\right.
 \end{equation}
 The parameter $B_1$ seems to disappear from the system, but in fact it is hidden in the additional relation $(Z-\mu)^2+P^2=B_1^2$.
 The first two equations of (\ref{4D1}) are very similar to the ones from the system  investigated by De Maesschalck et al. (after the transformation). The parameter $\omega$ in the paper of Shimizu et al. (2011) is a main bifurcation parameter and based on that   results   we may assume that it belongs to the interval $(0,1)$. Numerical results indicate that fixing $\omega$ and taking $\mu$  as a  bifurcation parameter   is possible and moreover   we can take  $\omega$ small enough so the system (\ref{4D1}) can be treated as a three time scale system. To be more precise we assume $\omega=O(\epsilon)$.
 
 What is clearly  completely different from the systems (\ref{KP}) and (\ref{jap}) is the return mechanism: in the system (\ref{KP}) $w$ grows as long as $\mu$ is positive and $z$ is small enough. In (\ref{4D1}) both $Z$ and $P$ are in fact periodic functions. The domain of $P$ is the interval $[-B_1,B_1]$, while the domain of the variable $Z$ depends on the parameter $\mu$: $Z\in[-B_1+\mu, B_1+\mu]$. Geometrically speaking we observe the trajectories "reflecting" from the ends of the domain. No change of direction with respect to $Z$ or $P$ is possible outside the "walls" of the domain. Another difference is a lack of stationary points for (\ref{4D1}). If we solve the equations for a stationary points  we obtain that $P=0$ together with $Z=\mu$ which imply $B_1=0$ and no perturbation. Instead of a singular point we obtain a periodic solution. For the same reason there is no Hopf bifurcation as such.

\begin{figure}
\centering
\subfigure[]
    {\includegraph[width=0.4\textwidth, height=0.25\textheight]{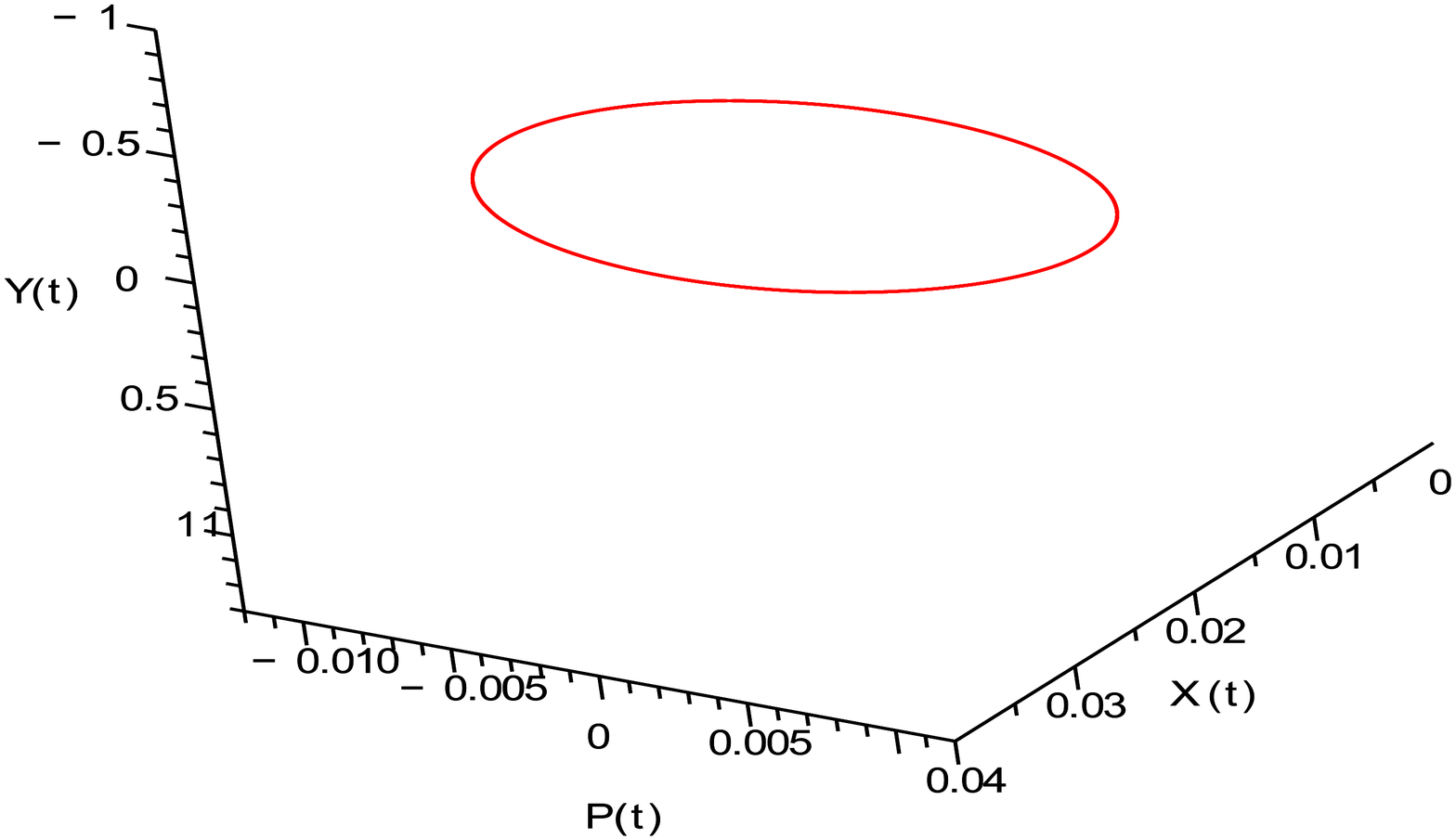}\label{little}}
\subfigure[]
    {\includegraph[width=0.4\textwidth, height=0.25\textheight]{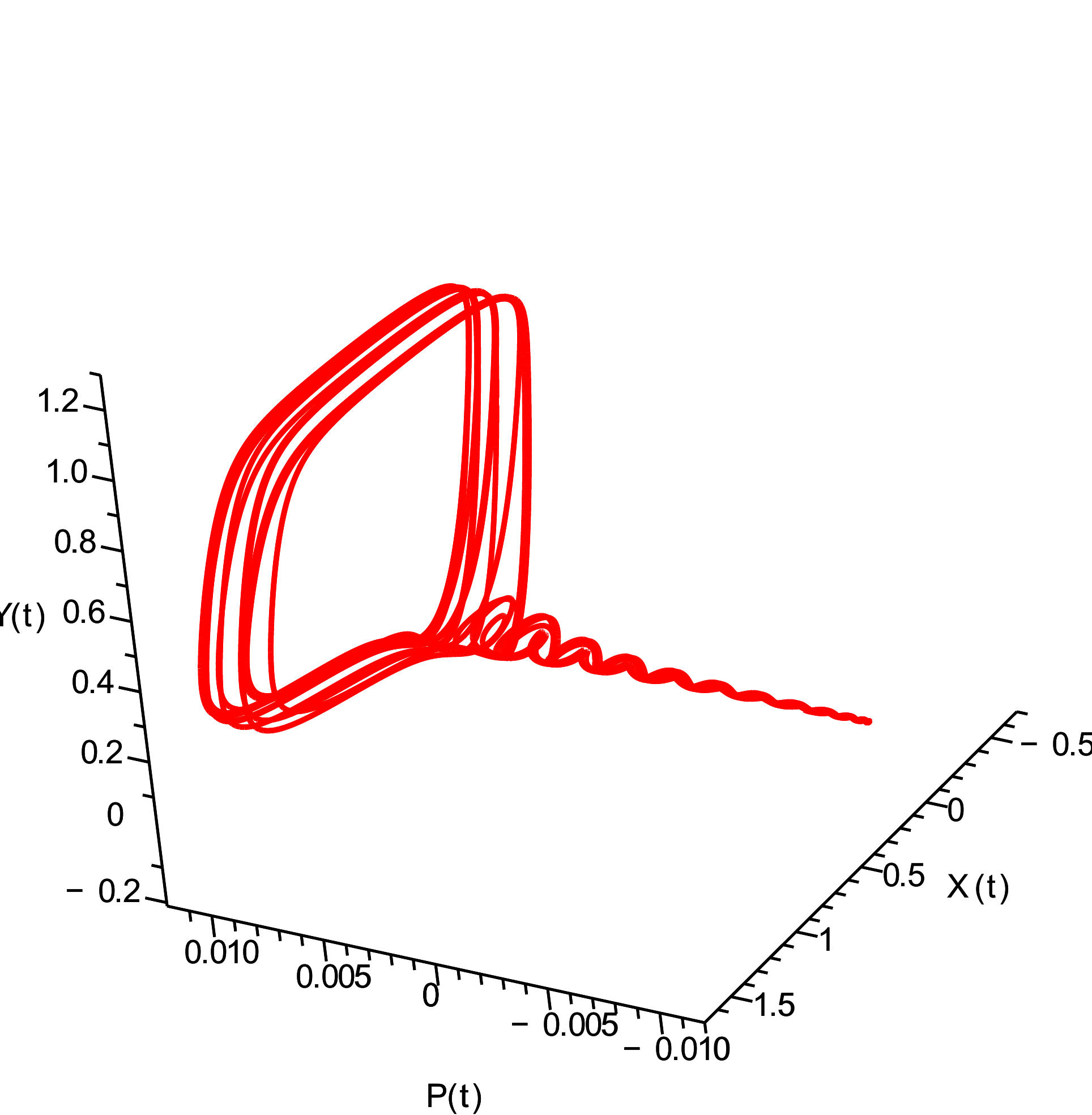}
        \label{needle}}
        \subfigure[]
    {\includegraph[width=0.4\textwidth, height=0.25\textheight]{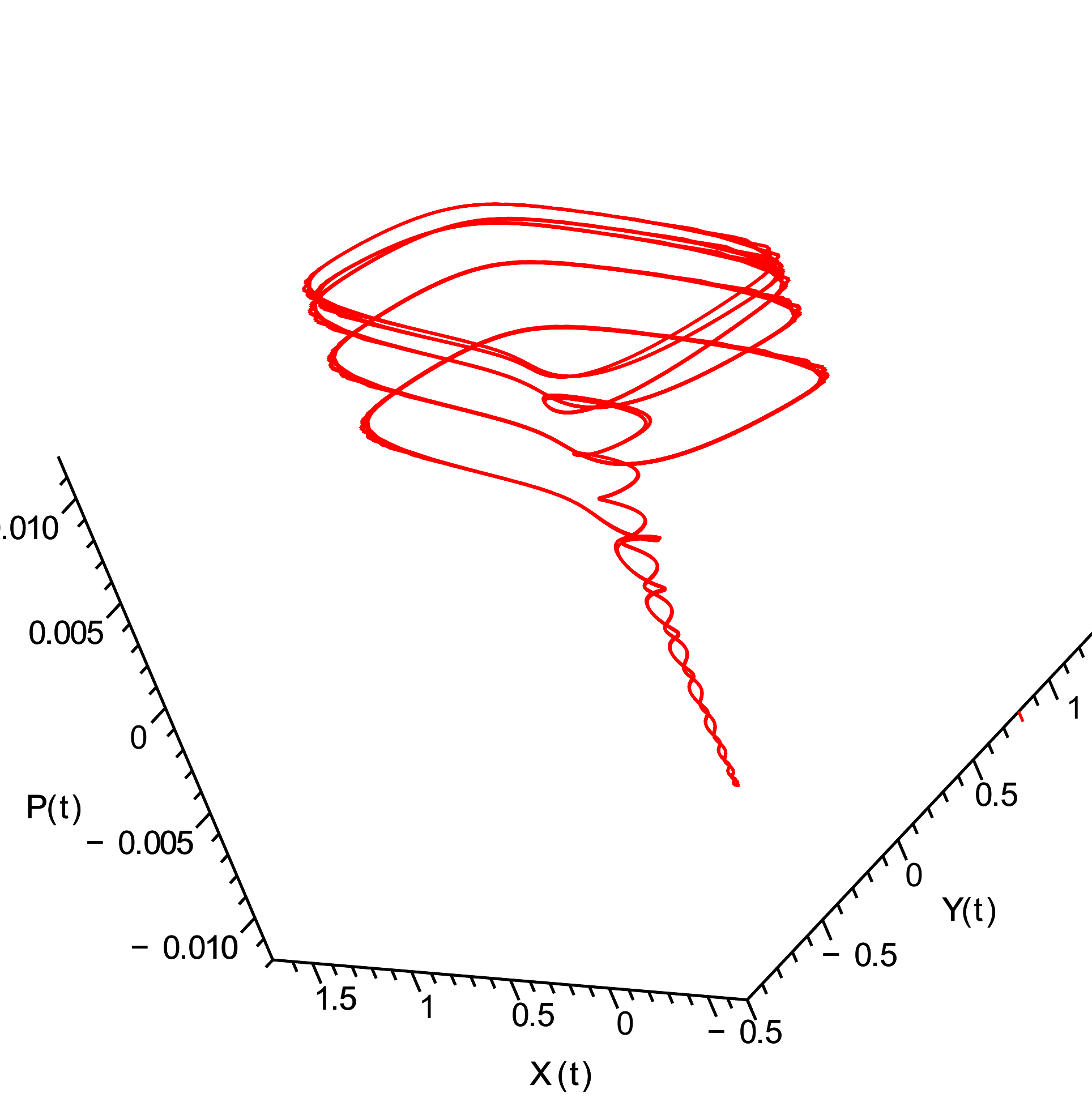}
        \label{205_3d}}
        \subfigure[]
    {\includegraph[width=0.4\textwidth, height=0.25\textheight]{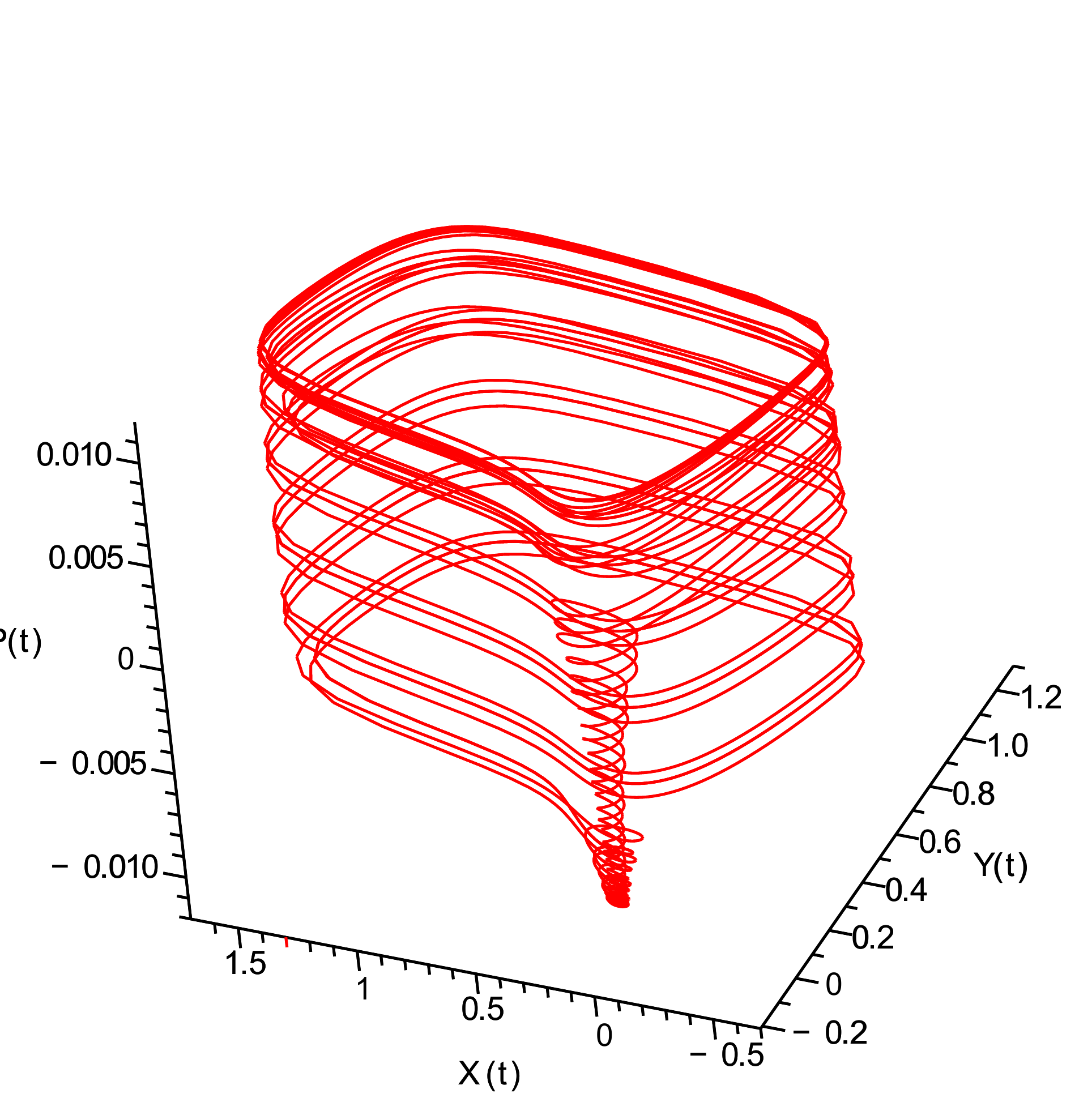}
        \label{203}}
        \subfigure[]
    {\includegraph[width=0.4\textwidth, height=0.25\textheight]{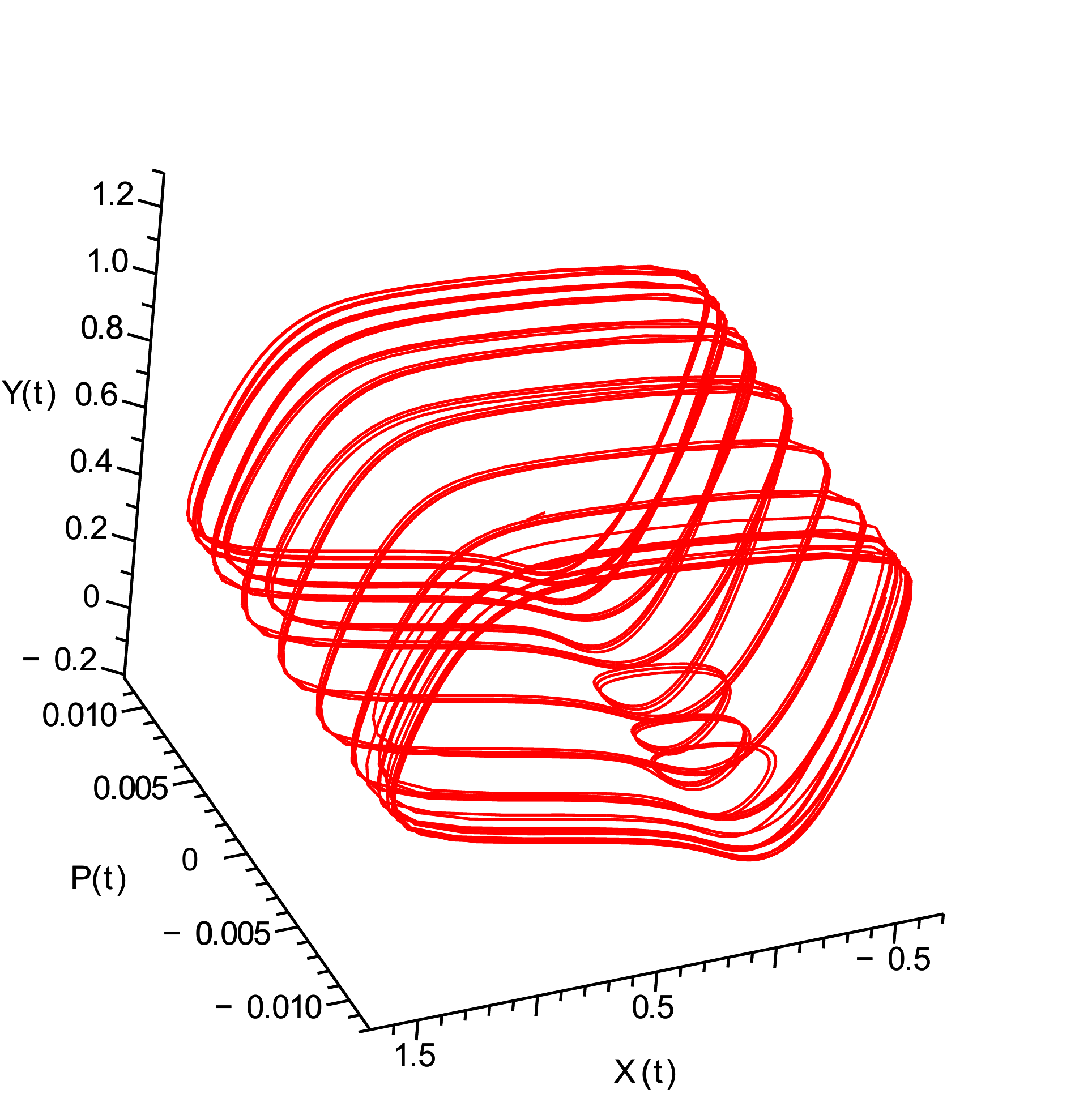}
        \label{201}}
         \subfigure[]
    {\includegraph[width=0.4\textwidth, height=0.25\textheight]{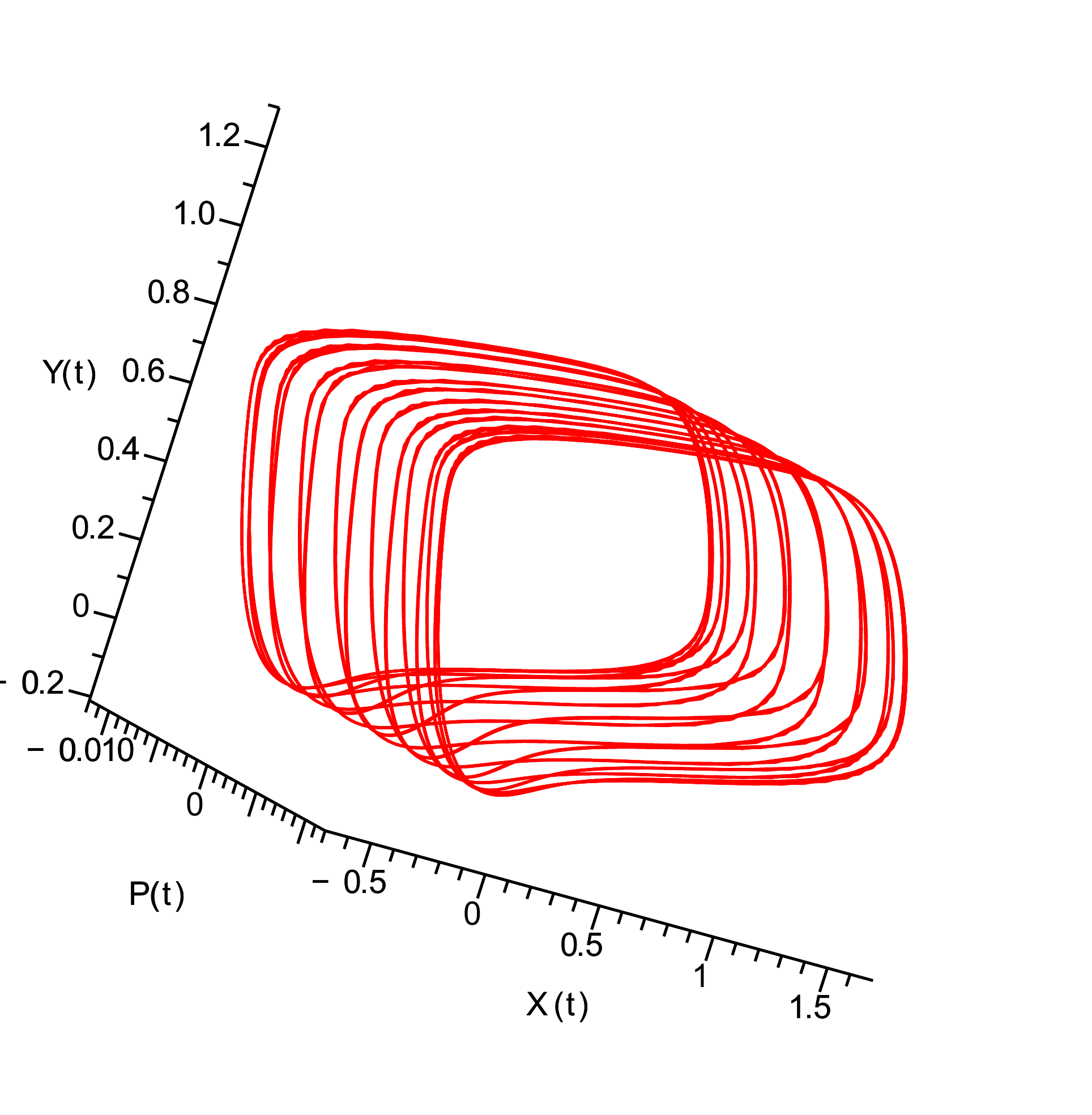}
        \label{200}}

\caption{Evolution from a small cycle to  relaxation oscillations for changing $B_0$. $\epsilon=\omega=B_1=0.1$, $k_1=0.9$ and (a) $B_0=0.212$, (b) $B_0=0.206$, (c) $B_0=0.205$, (d) $B_0=0.203$,  (e) $B_0=0.201$, (f) $B_0=0.200$. }\label{evolution}   
    \end{figure}

  On the figure \ref{evolution} we present some typical solutions of (\ref{4D1}) for $\epsilon=\omega=0.1$, $B_1=0.01$, $k_1=0.9$ and changing $B_0$.
Note that  $\mu=0$ corresponds to $B_0\approx 0.231$.  On the figure \ref{little} we see the starting periodic solution with a small amplitude in $X$ and $Y$, it corresponds to a stable stationary point in (\ref{KP}). With the growing $|\mu|$ this attracting  periodic orbit disappears and we observe attracting MMOs. On the next pictures we observe how the shape of mixed mode oscillations are changing with $B_0$. First small oscillations after reflection make almost the same way back (figure \ref{205_3d}), then the change from small oscillations to large happen earlier and earlier
 until on the figure \ref{201} we observe no more reflection, just several small oscillations before the jump. This situations correspond to the one studied by Krupa et al. (2008). In this paper there is an assumption, that the number of small oscillation should be not too big, otherwise the approximations are not valid anymore. 
 On the last figure we can see a trajectory corresponding to a relaxation oscillation in  the paper of Krupa et al. (2008). It was shown in the paper of Szmolyan and Wechselberger (2004) that this trajectory lies on the torus.
 \begin{figure}\centering
\includegraph[width=0.4\textwidth, height=0.25\textheight]{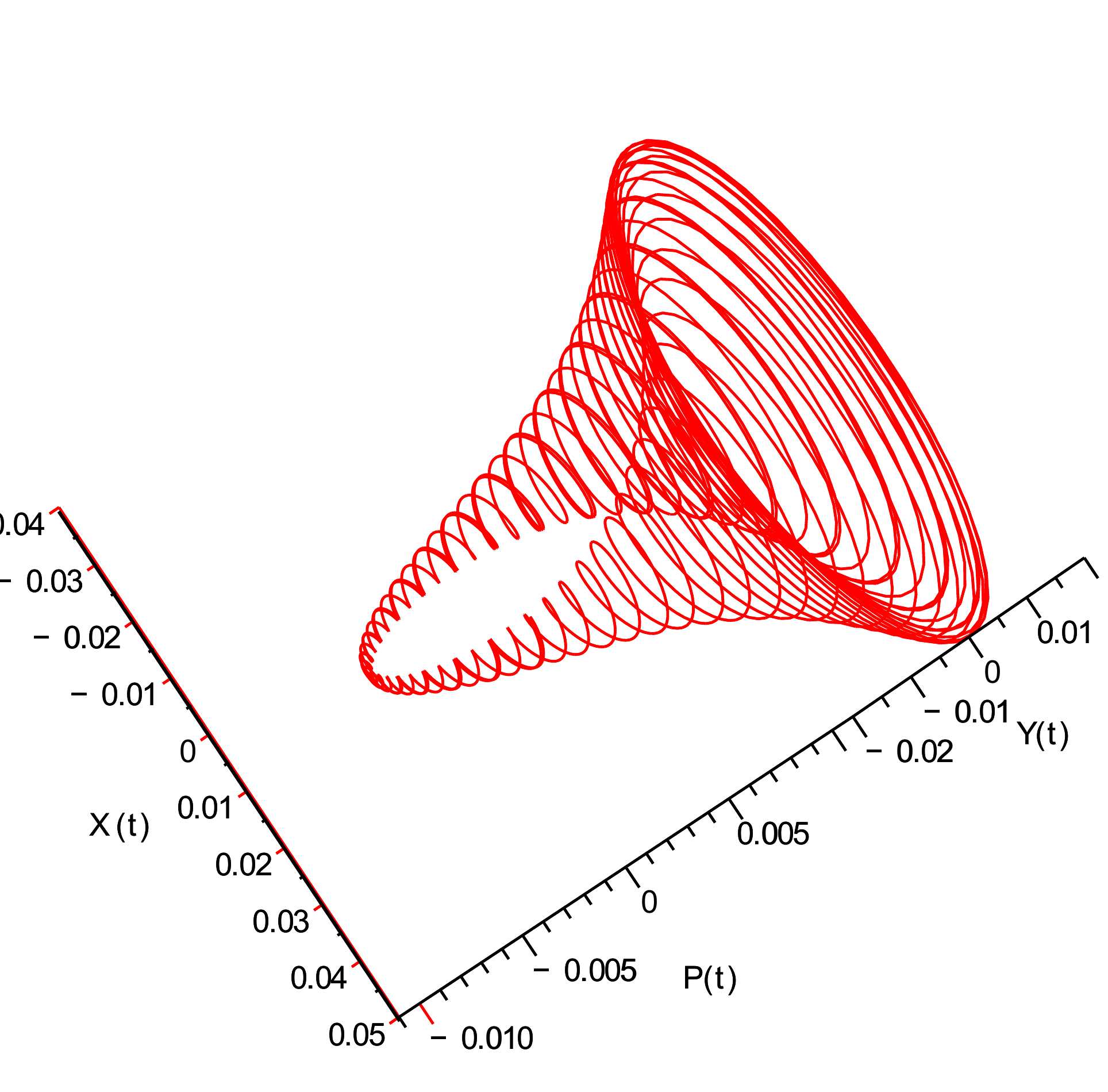}
\caption{Small torus. $\epsilon=\omega=B_1=0.1$, $k_1=0.2$, $B_0=0.4945$, $\mu\approx -0.0058$.}\label{torus}   
    \end{figure}
    
 These pictures  raise a natural question what is happening between the small cycle and the first observed MMO. In (\ref{KP}) there exist a parameter range where we may observe a small stable cycle born in the singular Hopf bifurcation  (see Guckenheimer 2008),  this cycle grows and  undergoes multiple  series of  the period doubling cascades which lead to MMOs.  In the paper by Petrov et al. (1992) one can find a bifurcation diagram with so-called isolas corresponding to different patterns of MMOs. In case of the system (\ref{4D1}) instead of a small cycle we would expect a small torus. Near $k_1=0.9$ such a solution is difficult to detect. The reason is that as we stated before the planar Hopf bifurcation is subcritical for this value of $k_1$. Thus planar small cycles are unstable and in the full system the possible tori most probably have saddle properties.
 So to prove our hypothesis  numerically we change the value of $k_1$ to $0.2$ (which corresponds to a supercritical Hopf bifurcation) and indeed for $B_0=0.4945$ ($\mu\approx -0.0058$ ) we observe the small torus presented on figure\ref{torus}. \\

     \begin{figure}\centering
\includegraph[width=0.4\textwidth, height=0.25\textheight]{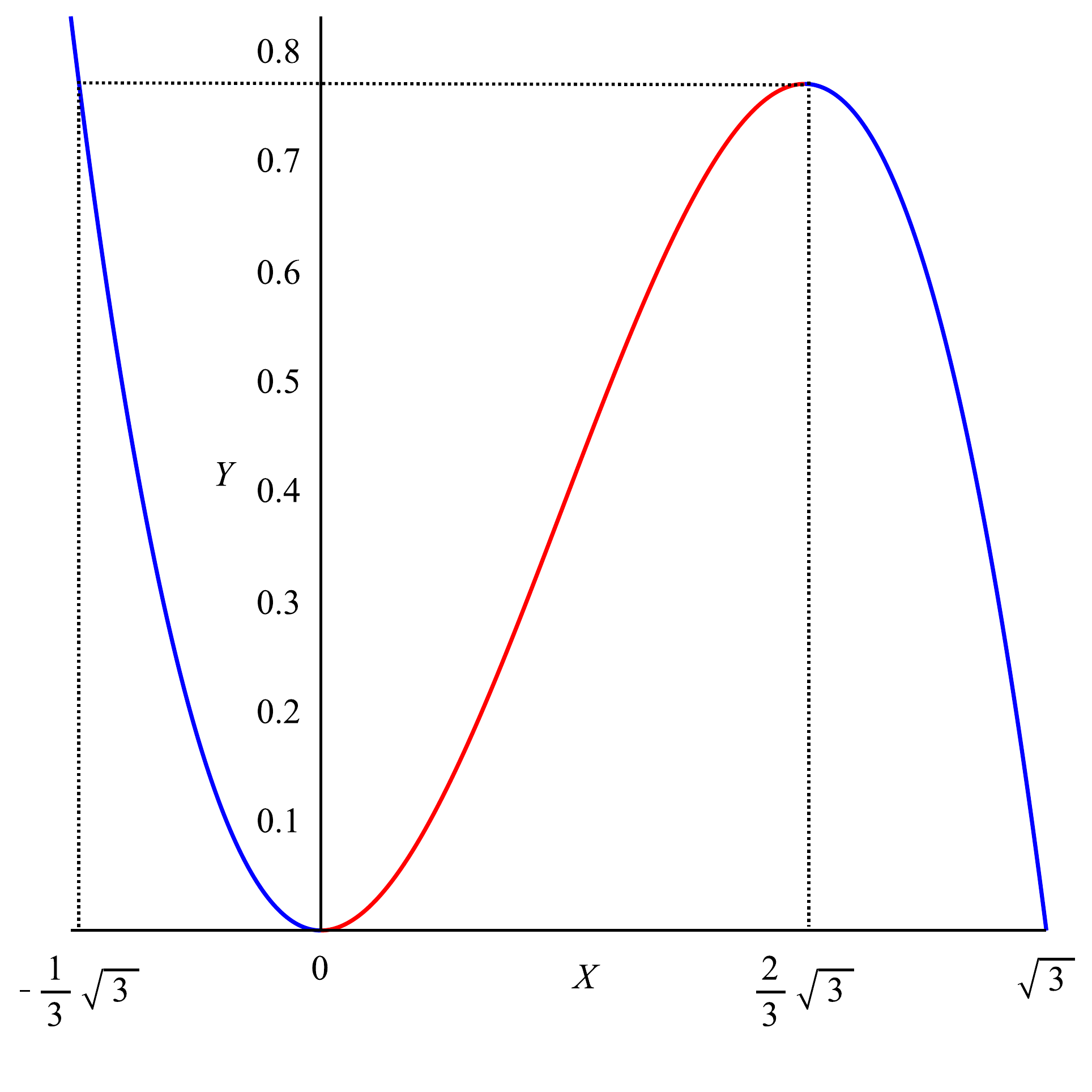}
       \caption{Slow manifold of the system (\ref{4D1}). Blue and red lines represent respectively attractive and repelling parts.}\label{sm}   
    \end{figure}
    Let us now try to understand the mechanisms which form mixed mode oscillations with small number of SAOs. On the figure \ref{sm} we present a planar projection of a slow manifold for the  unperturbed system (\ref{4D1}) ($\epsilon=0$). It consists of left and right attracting parts ($S_0^{a-}$ and $S_0^{a+}$) and in the middle there is a repelling part ($S^r_0$). When some perturbation is present ($\epsilon>0$) these three parts perturb respectively into $S_{\epsilon}^{a-}$, $S_{\epsilon}^{a+}$ and $S_{\epsilon}^{r}$. Outside the fold regions the perturbed manifolds "inherits" the dynamics from the unperturbed system following the rules of Fenichel's theory. The neighborhood  of the fold requires special analysis: we will show the presence of fold singularities and use a rescaling of the variables to understand the dynamics near $Z=0$. 
     
 In the paper by Krupa et al. (2008) (among other results) there is presented a very detailed analysis of the MMO's of the type $1^k$ (one large oscillation followed by $k$ small). The intervals of the parameter such as $1^k$ solution exist and is attracting are approximated. It is done by constructing a two-dimensional  Poincar\'{e} map defined on the plane $v=0$. This plane contains the fold and intersects an unperturbed slow manifold transversally. Poincare\'{e} map is divided into four parts:  fold area, return mechanism and transitions between fold area and return mechanism. The key simplification of the problem is the fact that the two dimensional map can be successfully approximated by a one dimensional map. The reason is that each trajectory jumping off the left fold is getting attracted to the right attracting part of the slow manifold ($S^{a+}_{\epsilon}$), then following the perturbed flow reaches the right fold region, then jumps off again to follow a fast flow back to the left attracting branch ($S^{a-}_{\epsilon}$). Again, following the slow flow the trajectory gets exponentially close to the straight line $C^{-}_{\epsilon}$ which denote the intersection of $S^{a+}_{\epsilon}$ and the plane $v=0$. As a result it is possible to study a one dimensional Poincar\'{e} map $C^{-}_{\epsilon}\to C^{-}_{\epsilon}$, which is a much easier task. The line $C^{-}_{\epsilon}$ contains a critical canard point which corresponds to the transversal intersection of the left attracting manifold $S_{\epsilon}^{a-}$ and the repelling manifold $S_{\epsilon}^{r}$. Every trajectory starting from the point on $C^{-}_{\epsilon}$ with $w$ coordinate smaller that $w^{cr}$ undergoes small oscillations, while if it starts on the other side it jumps to $S_{\epsilon}^{a+}$ as described before following the fast flow first and then creating a large loop (we call it return mechanism). It is possible to define sectors of rotation $RS^k$ - intervals  of  $C^{-}_{\epsilon}$ such as if the trajectory starts in $RS^k$ it undergoes exactly $k$ small oscillations before leaving the fold region. Precise analysis of the dynamics near the fold is only possible for $w$ small enough. In that case the solutions can be approximated as perturbations of a certain integrable two dimensional system.

Our numerical and analytical observations indicate that it should be possible to perform an analogous reduction for the system (\ref{4D1}). As it was mentioned before our current goal is just  to indicate some similarities and differences as well as prepare a background for a further detailed analysis. 
 
 \section{Folded singularities}
Let us start our analysis with the examination of the folded singularities of the system (\ref{4D1}).
In order to do so we project system (\ref{4D1}) on the slow manifold $Y=\sqrt{3}X^2-X^3=f(X)$. De-singularized (see Desroches et al. 2012) system has the following form:
 \begin{equation}\label{4D1pr}
        \left\{
        \begin{array}{ll}
        X'=-X-k_1\,f(X)+Z\\
        P'=\omega f'(X)\, (\mu-Z)\\
        Z'=\omega f'(X)\, P.
        \end{array}\right.
 \end{equation}
 Here $()'$ in the RHS denotes the rescaled time-derivative $(\dot )f'(X)$. Time flow changes the directory for $X$ such as $f'(X)<0$, so on the attracting branches of the manifold.
 
Solving the equations for the stationary points we obtain  $(0,\pm\sqrt{B_1^2-\mu^2},0)$ on the left fold and $(\frac{2}{3}\sqrt{3},P_0,\frac{2}{3}\sqrt{3}\,(\frac{2}{3}k_1-1))$ on the right fold. $P_0$ denotes a value of $P$ corresponding to $Z=\frac{2}{3}\sqrt{3}\,(\frac{2}{3}k_1-1)$. For the considered values of the parameters: $k_1\in(0,1)$, $B_1$ and $\mu$ small there are no singularities on the right fold.  Assuming $|\mu|<B_1>0$ we have two stationary points $A_1=(0,\sqrt{B_1^2-\mu^2},0)$ and $A_2=(0,-\sqrt{B_1^2-\mu^2},0)$ on the left fold. Corresponding eigenvalues are: 
$$\lambda_{1,2,3}(A_1)=\left( 0,\frac{1}{2}\pm\frac{1}{2}\sqrt{1-8\sqrt{3}\omega\sqrt{B_1^2-\mu^2}}\right)$$ and $$\lambda_{1,2,3}(A_2)=\left(0,\frac{1}{2}\pm\frac{1}{2}\sqrt{1+8\sqrt{3}\omega\sqrt{B_1^2-\mu^2}}\right).$$ 
In both points zero eigenvalue corresponds to $P$-direction. Assuming $B_1$ and $\mu$ small enough  the first point is a node while the second one is a saddle in the $(X,Z)$ space. For $\mu=\pm B_1$ or $\omega\to 0$ two singular points melt into a single one placed in the origin and it is a  folded saddle-node (in $(X,Z)$ space again) with  the eigenvector corresponding to zero eigenvalue transversal to a fold. In case of the system (\ref{KP}) we have only one singular point and the saddle-node appears as folded saddle bifurcating into a folded node. However in the analyzed range of parameters singularity moves outside the assumed domain ($w\sim O(\epsilon)$). The same situation is present in the system (\ref{4D1}) for $|\mu|>B_1$ (which seems to be satisfied for  the parameter range where we observe MMOs with a small number of SAOs).  Krupa and Wechselberger (2010)  investigate the folded saddle-node singularity in the domain wide enough to take into account the singularities but in that case different phenomena play a key role.


 \section{Fold region in the rescaled coordinates}
 In this section we investigate the fold area using a classical approach of geometrical theory singular perturbations. We make a rescaling which is in fact identical to considering a family chart of blown-up coordinates:
  \begin{equation}X=\epsilon^\frac{1}{2} \bar{X},\qquad Y=\epsilon \bar{Y},\qquad P=\epsilon^\frac{1}{2} \bar{P}, \qquad Z=\epsilon^\frac{1}{2} \bar{Z}. \end{equation} After an additional time rescaling (we keep the $\dot{()}$ notation) the system (\ref{4D1}) takes form
  \begin{equation}\label{blow}
        \left\{
        \begin{array}{ll}
        \dot{\bar{X}}=-\bar{Y}+\sqrt{3}\bar{X}^2-\sqrt{\epsilon}\bar{X}^3\\
        \dot{\bar{Y}}=\bar{X}-k_1\,\sqrt{\epsilon}\bar{Y}+\bar{Z}\\
        \dot{\bar{P}}=\omega\, (\bar{\mu}-\sqrt{\epsilon}\bar{Z})\\
        \dot{\bar{Z}}=\omega\, \sqrt{\epsilon} \bar{P}.
        \end{array}\right.
 \end{equation}
  For $\epsilon=0$ and $\bar{Z}=0$ (which is satisfied for $B_1$ and $\mu=0$) first two equations of the system (\ref{blow}) form an integrable system with a constant of motion given by
  \[H( \bar{X},\bar{Y})=\frac{1}{2}e^{-2\sqrt{3}\,\bar{Y}}\left(-\bar{X}^2+\frac{\bar{Y}}{\sqrt{3}}+\frac{1}{6}\right).\] 
  The special solution corresponding to $H(\bar{X},\bar{Y})=0$
\[\bar{\gamma}^0_0(t)=\left(\frac{1}{2\sqrt{3}}t,\frac{1}{4\sqrt{3}}t^2-\frac{1}{2\sqrt{3}}\right)\] separates the area of periodic solutions  and open level curves. This solution also corresponds to a singular canard (singular limit of  intersection of the manifolds $S_{\epsilon}^{a-}$ and $S_{\epsilon}^r$).

  The integrable system is exactly the same as the one examined by De Maesschalck et al. (preprint). Moreover, if we assume $\omega=O(\epsilon)$ as mentioned before, we can treat the full four-dimensional system as a perturbation of the integrable two dimensional one. The singular canard trajectory perturbs into a strong canard lying on the intersection of perturbed attractive and repelling manifolds.  To compute the corresponding value of $\bar{Z}$ (we call it $\bar{Z}_{cr}$) we may follow the calculations from the paper by De Maesschalck et al. (the only difference is a plus sign at $\bar{Z}$ in the second equation) and conclude   that the critical value of $\bar{Z}$ is given by 
  \[\bar{Z}_{cr}=-\frac{\sqrt{3}(1+2k_1)}{24}\sqrt{\epsilon}. \] 
  The idea behind this formula is as following: $\bar{Z}_{cr}$ divides the area of almost-periodic orbits from non-periodic ones. Thus the orbit passing this critical point can be treated as a periodic one, but with an infinite period time. Small $\omega$ allows us to treat $\bar{P}$ and $\bar{Z}$ as constants, while small $\epsilon$ guarantees that the orbits could be approximated by the orbits of the integrable planar system. The computations  are almost identical to those made by De Maesschalck et al., but their presentation would involve a lot of additional notations. \\

If we recall the original variables we obtain $Z_{cr}=-\frac{\sqrt{3}(1+2k_1)}{24}\epsilon$. Since there is a relationship between $P$ and $Z$ we can compute $P_{cr}=\pm\sqrt{B_1^2-(Z_{cr}-\mu)^2}$. That means that there exist two canard trajectories (obviously if $\epsilon$ and $\mu$ are small enough both $Z_{cr}$ and $P_{cr}$ belong to the domain). We can observe this phenomena for instance on the figure \ref{201}: the area of jumps is followed by the small oscillations and then by jumps again. However in the system (\ref{4D1}) this double canard occur due to the symmetry of the trigonometric functions, it could be interesting to construct a three-dimensional polynomial system with similar geometry.
  \section{Return mechanism}
To obtain an approximation of the return mechanism let us go back to the projection on the slow manifold described by the system  (\ref{4D1pr}). For $\epsilon$ and $\omega$ small enough we may assume that the change of $P$ during "fast time" intervals can be neglected, so the slow manifold projection provides us enough information to approximate  the map.
If we divide the second equation of  (\ref{4D1pr}) by the first one we obtain
\begin{equation}\label{dpdx} \frac{dP}{dX}=\omega\frac{f'(X)(\mu-Z)}{X-k_1\,f(X)+Z}.\end{equation}
We assume that $Z$ is small comparing to $X-k_1\,f(X)$ which is justified by the domain where $Z$ is defined. Then (recalling the relationship $(Z-\mu)^2+P^2=B_1^2$) we may rewrite (\ref{dpdx}) as
\begin{equation}\label{dpdx1} \frac{dP}{\pm\sqrt{B_1^2-P^2}}=\omega\frac{f'(X)dX}{X-k_1\,f(X)},\end{equation}
where the sign at the square root depends on the direction the trajectory is following. It is $+$ for increasing $P$ and $-$ for decreasing. 
Let us introduce an auxiliary variable $W=\omega\, t$. To cover all possible situations it is sufficient to take $W\in[0,3\pi]$.  We can easily compute that    
 $dW=\frac{dP}{\pm\sqrt{B_1^2-P^2}}$, which transforms (\ref{dpdx1}) into a simpler dependence
\begin{equation}\label{dwdx}dW=\omega\frac{f'(X)dX}{X-k_1\,f(X)}.\end{equation}
It is clear from the figure \ref{sm} that for the full return loop the variable $X$ needs to run from $\sqrt{3}$ to $\frac{2}{3}\sqrt{3}$ (right attractive manifold) and then from $-\frac{\sqrt{3}}{3}$ to $0$ (left one). 

If we denote by $P_0$ a start point on $S_{\epsilon}^{a-}$ close to a jump area then a corresponding $W_0$ is assumed to be  given by $\arccos\frac{P_0}{B_1}$ if in $P_0$ $P$ is a decreasing function of time and $2\pi-\arccos\frac{P}{B_1}$ otherwise.
 We may approximate the result of the return mechanism as
\begin{equation}\label{return}W_1=W_0+\omega \left( \int_{\sqrt{3}}^{\frac{2}{3}\sqrt{3}}\frac{f'(X)dX}{X-k_1\,f(X)}+ \int_{-\frac{\sqrt{3}}{3}}^{0}\frac{f'(X)dX}{X-k_1\,f(X)}\right)= \end{equation}
\[=W_0+3\omega(1-\frac{k_1}{2}).\]
 Now, depending in which part of the domain we landed we have
\[  W_1^-=\arccos\frac{P_1}{B_1},\quad W_1^+=2\pi-\arccos\frac{P_1}{B_1}\quad \textrm{or} \quad W_1^==2\pi+\arccos\frac{P_1}{B_1}.\]
Checking all possible combinations (the trajectory can come back without or with hitting the domain boarder) we obtain the formula

\[ P_1=B_1\cos \,(\arccos\frac{P_0}{B_1}\pm 3\omega (1-\frac{k_1}{2})),\]
 where the sign is $-$ if we follow the direction of increasing $P$ and $+$ otherwise.
Numerical experiments confirm this approximation. Let us underline that this formula covers also the case when there is a "reflexion" effect in the way. The trigonometric functions are responsible for the "compression" close to the border of the domain while in the paper by Krupa et al. (2008) the distance between initial and end points of the return mechanism is approximately constant.

\begin{figure}
\centering
 \includegraph[width=1\textwidth, height=0.25\textheight]{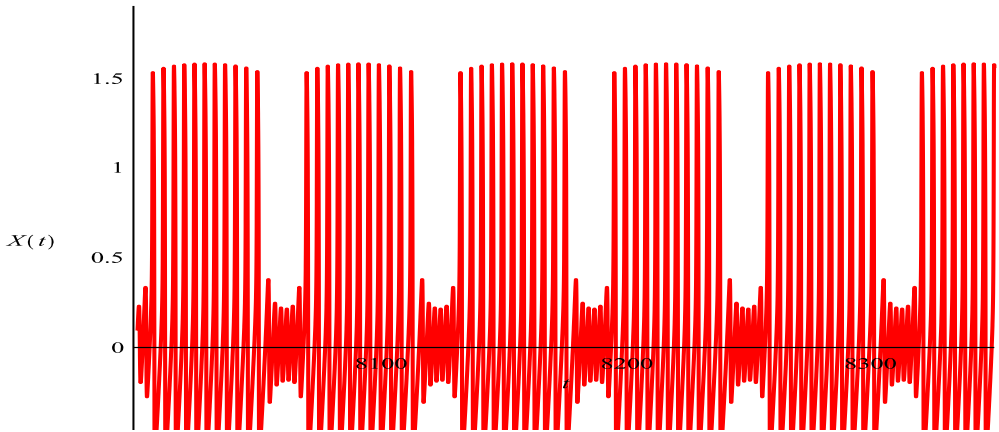} 
\caption{$\epsilon=\omega=0.1$, $k_1=0.2$, $B_1=0.01$, $B_0=0.4815$, $\mu\approx -0.0119$.}\label{period}   
    \end{figure}

The formula (\ref{return}) shows that the number of LAOs per length of the domain depends only on the parameters $k_1$ and $\omega$. This approximation is good enough for the Poincar\'{e} map, but unfortunately does not allow us to generate periodic orbits. It is however practically easier to catch a (numerically) periodic orbit for $k_1=0.2$ then $k_1=0.9$, one of the possible reason is a clearer picture due to less LAOs per period. An example of such an orbit is presented on the figure \ref{period}.

\section{Discussion}

In this paper we presented a locally three dimensional system which is equivalent to a Bonhoeffer--van der Pol model. Keeping the parameter region close to the examples from the paper of Shimizu et al. (2011) we discussed the similarity of that system to  known models with folded saddle-node II singularity. Some additional assumption were made in order to separate the time scales and eliminate the effect  of the stationary points. A brief analysis of the fold region and an approximation of the return mechanism were provided.

We observed several differences such as a presence of two canards per period and non-homogeneous distribution of the return mechanism. Basically, these effects are present due to the choice of the variables. On the other hand this choice not only draw a parallel to the known  and well analyzed systems, but also could be useful for the construction of polynomial systems with interesting dynamics.\\

 Further analysis of the system could be especially interesting if we recall that the original system (\ref{fvdp}) models a simple circuit, which can be use for experiments with the parameters. That implies, that this circuit can be a very simple physical model for the three time scale systems with a folded saddle node singularity.
 
\begin{figure}
\centering
\subfigure[]
    {\includegraph[width=0.4\textwidth, height=0.25\textheight]{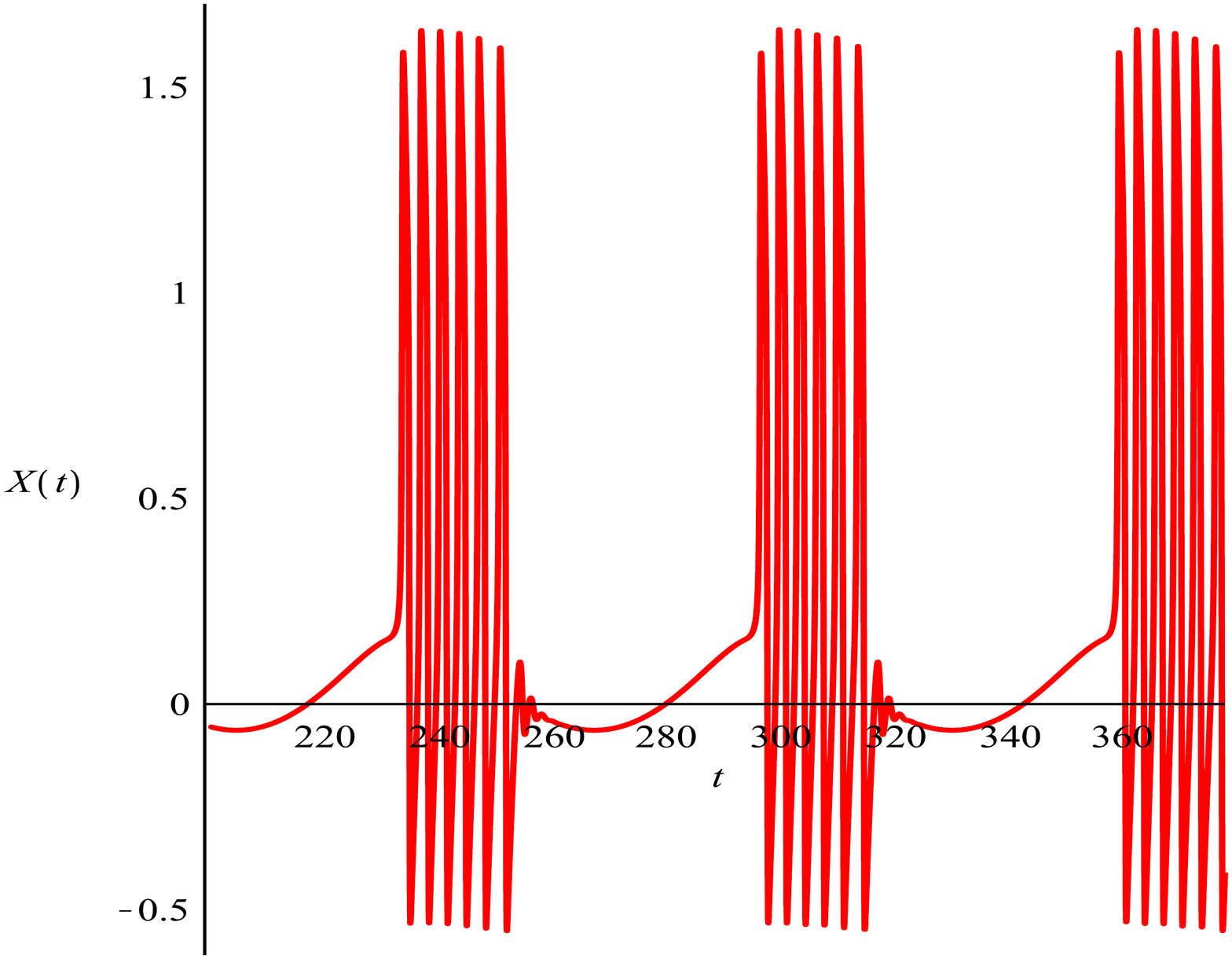} \label{burst_2d}}
\subfigure[]
    {\includegraph[width=0.4\textwidth, height=0.25\textheight]{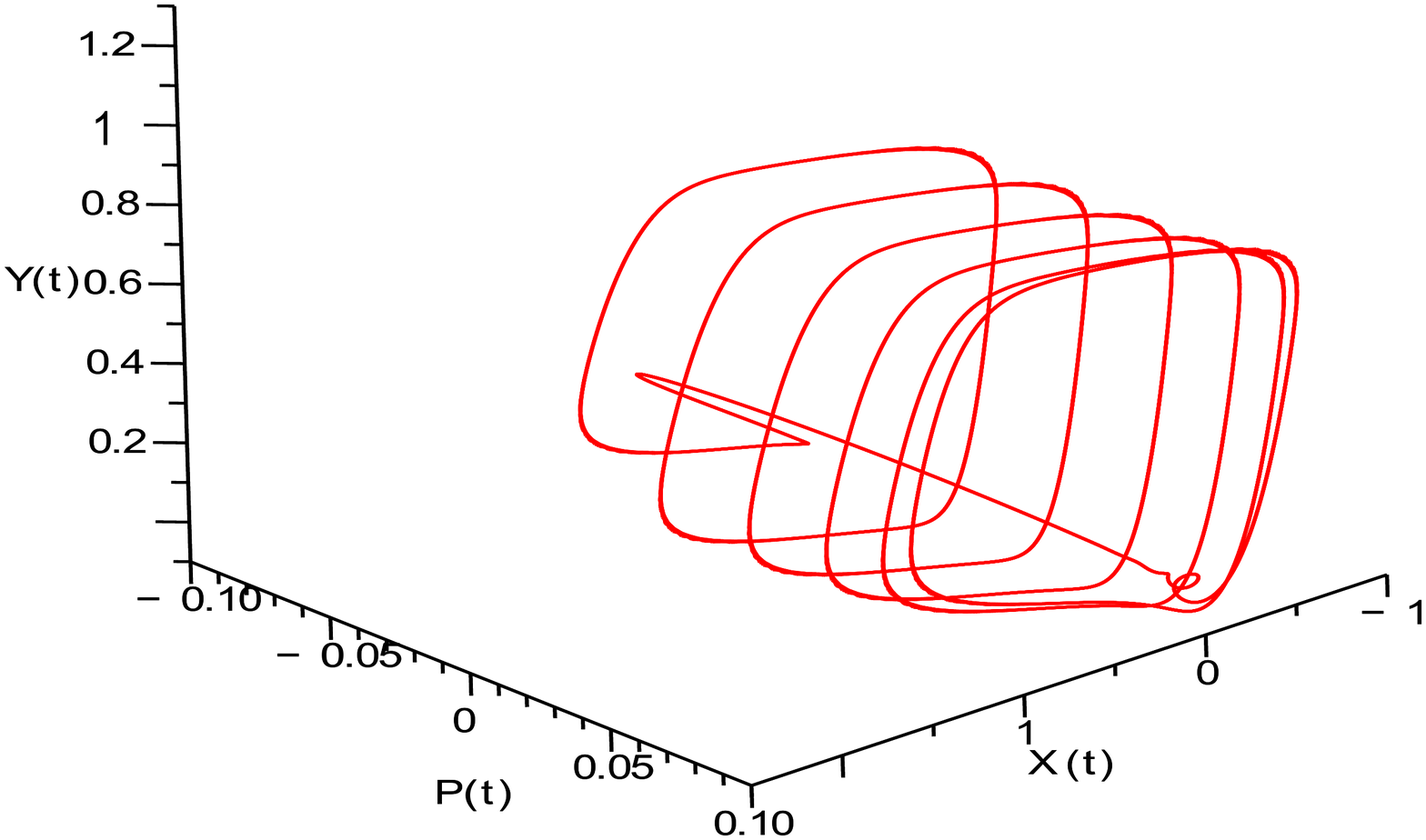}
        \label{burst_3d}}
    \caption{$\epsilon=\omega=B_1=0.1$, $k_1=0.9$,  $B_0=0.201$, $\mu\approx -0.03$.}\label{burst}   
    \end{figure}

In our research we concentrated on the indication of the parameter range where the particular type of dynamics takes place. It could be interesting to go outside this region and take a look on the nearby bifurcations based on  the papers of Guckenheimer et al. (2003)  and Bold et al. (2003). To underline how different the behavior of the trajectory  could be, on the figure \ref{burst} we present the periodic MMO with so-called bursting phenomena present. This type of solution is typical for the neuronal models (see Izhikevich 2007). The parameters are like in case of the figure \ref{201}, except $B_1=0.1$. This change also cause that the stationary points on the fold exist even for relatively large $|\mu|$. Another interesting phenomenon to study is   presence of the delayed  Hopf bifurcation (see Krupa and Wechselberger 2010, Neishtadt 1987).

\section*{Acknowledgments} Author is very grateful to prof. Peter De Maesschalck for many stimulating discussions.\\
 The presented research was supported by the Research Foundation Flanders (FWO) under grant number
G.0939.10N and by Polish Ministry of Science and Higher Education.

\end{document}